\documentclass[doublecol]{epl2} 
\usepackage{dcolumn}
\usepackage{bm}
\usepackage{xcolor}
\usepackage[dvips]{epsfig}
\usepackage{graphicx} \usepackage{amsmath} \usepackage{amssymb}

\newcommand{\comment}[1]{}

\newcommand{\BEA}{\begin{eqnarray}}
\newcommand{\EEA}{\end{eqnarray}}

\newcommand{\bq}{\begin{equation}}
\newcommand{\eq}{\end{equation}}
\newcommand{\be}{\begin{eqnarray}}
\newcommand{\ee}{\end{eqnarray}}
\newcommand{\ba}{\begin{align}}
\newcommand{\ea}{\end{align}}

\newcommand{\A}{{\cal A} }
\newcommand{\B}{{\cal B} }

\renewcommand{\A}{\mathbb{A}}
\renewcommand{\B}{\mathbb{B}}
\renewcommand{\H}{{\cal H}}

\title{Common causes for quantum identical particles}

\shorttitle{Common causes for quantum identical particles}

\author{Arshak Hovhannisyan\inst{1} \and Stefan Weigert\inst{2} 
\and Armen E. Allahverdyan\inst{1}}
\shortauthor{A. Hovhannisyan \etal}

\institute{                    
  \inst{1} Alikhanyan National Laboratory - Yerevan 0036, Armenia\\
  \inst{2} University of York, Department of Mathematics - Heslington, York YO10 5GH, UK
}


\abstract{ Violations of Bell's inequalities imply that joint probabilities generated by non-commutative measurements on two (non-identical) quantum particles do not have a single common cause. But joint probabilities generated for such non-identical particles via commutative measurements do have non-trivial common cause variables. We focus on commutative measurements and consider two identical quantum particles, whose density matrices and observables (hermitian operators) are necessarily permutation-symmetric. It is natural to demand that the common cause describing joint probabilities is also permutation symmetric, i.e., it acts symmetrically on both particles. Looking at various ways of defining joint probabilities from the same measurement data, we conclude that either symmetric common causes need not exist (i.e., that the particles can be hiddenly distinguishable), or that symmetric screening variables exist, but they are trivial, i.e., no single common cause can explain all single-measurement correlations. 
 }

\begin{document}

\maketitle
\section{Introduction}
Probabilistic causal inference is based on the principle of common cause (PCC) that seeks a conditionally independent representation of the probability $p(A,B)$ of two independent random variables $A$ and $B$ \cite{reich,suppes,uffink,szabo}. The common cause is a random variable $C$ that leads to conditional independence $p(A,B|C)=p(A|C)p(B|C)$ of $A$ and $B$. PCC can be applied to explain $A$ and $B$'s dependency as a result of $C$, if two necessary conditions are met \cite{reich}:

{\it (i)} $A$ and $B$ do not influence each other; 

{\it (ii)} $C$ is in the common past of $A$ and $B$.

PCC was envisaged by Pearson and Yule in the early days of statistics \cite{historical}. It was developed by Reichenbach and Suppes \cite{reich,suppes}. PCC was generalized to the causal Markov condition, directed acyclic graphs, and the Markov blanket concept, pillars of the modern probabilistic causality \cite{pearl,janzing}. PCC led to pertinent implications in fields related to probabilistic inference: statistical physics (direction of time) \cite{reich,penrose1962direction}, cognitive psychology \cite{rehder2003causal}, evolutionary biology \cite{sober_common}, decision-making \cite{simpson}, {\it etc}. 

PCC got important applications in quantum mechanics; see Appendix and \cite{bell_cuisine,fraassen,wharton,szabo}. Consider two non-identical quantum subsystems $\A$ and $\B$ that interacted in the past, but are now far from each other to ensure the above applicability condition {\it (i)} of PCC. If the joint state $\rho$ of $\A+\B$ is non-entangled, then (by definition), all joint probabilities of random variables $A$ and $B$ (pertaining to $\A$ and $\B$, respectively) admit a single common cause $C$. However, there are entangled states, where several joint probabilities generated by noncommuting measurements do not have the same common cause $C$. This is the message of violating Bell's inequalities \cite{bell_cuisine,fraassen,wharton,szabo,berko,cushing,gudder1984reality,allahverdyan2018excluding}. Noncommuting measurements are executed on different $\A+\B$ (this explains why the same common causes need not exist \cite{theo}), but with the same state. Joint probabilities deduced from commutative measurements do have well-defined common causes, i.e., the generality of common causes is absent in quantum theory, not their existence as such; see Appendix. 

\comment{
Bell's inequalities and the underlying notion of quantum entanglement received enormous attention and sparked numerous discussions \cite{bell_cuisine,berko,cushing}. We do not enter into those discussions, but we still want to make 3 points. First, the non-existence of a single common cause led to reconsiderations of both conditions {\it (i)} and {\it (ii)}, i.e., to theories where distant subsystems can influence each other, or theories where causes are no longer located in the past \cite{wharton}. Second, the local hidden variable models (with which Bell's inequalities were initially formulated) are nothing but one specific mechanism with which one generates probabilistic common causes out of deterministic mechanisms subject to independent noises \cite{berko}. 
}

Here we apply PCC to probabilistic dependencies generated on (massive) identical particles, i.e., the particles that have the same values for all their time-independent parameters; e.g., mass, charges, Hilbert space dimension of spin, {\it etc}. We focus on probabilities generated via a single (i.e., commutative) measurement.
Recall that the status of quantum identical particles is quite different from their classical analogs \cite{landau2013quantum,ballentine}. Quantum particles are bosons or fermions that are described by permutation-symmetric density matrices, while physical observables (hermitean operators) acting on them are also necessarily permutation symmetric \cite{ballentine}. Otherwise, measuring a non-symmetric operator and selecting a definite result would make the post-measurement state non-symmetric. 

Given the permutation-symmetric status of identical particles, it is natural to demand that a common cause that describes two quantum identical particles be also permutation symmetric, i.e., it acts symmetrically on both particles, also for joint probabilities generated by commutative measurements. 

\comment{ 
Here we emphasize that due to the feature ${\it (i)}$ of PCC, we should look at the special class of local measurements, which are done by different (far away) particle coordinate detectors, each detector measuring (besides coordinates) a fixed, single-particle spin observable. }

Within the usual quantum formalism of identical particles, there appear to be at least two ways of defining joint probabilities from the results of commutative measurements. The first way is the standard one, and it implicitly involves labels for identical particles. The PCC applies to such joint probabilities, but there are cases in which the common cause cannot be symmetric. However, employing particle labels may not be the most constructive approach for understanding fundamental features of quantum identical particles \cite{mirman,krause1, krause2,we}. Hence, we also worked out an approach towards defining joint probabilities that explicitly avoids particle labels. Here, (symmetric) screening variables do exist, but they are trivial and do not amount to probabilistic causality: they provide no new information whatsoever, and they violate the anti-reflexive feature of causality: no variable can cause itself. 

These results can be reacted to in at least two different ways. First, we can assume that 
at least for certain states, the identical particles are still distinguishable at a hidden level; see \cite{we} for a similar assumption and some of its consequences. Then the usual probabilistic causality applies with nontrivial common causes that need not be symmetric with respect to particles, since they are distinguishable. Second, we can assume no single common cause random variable can (symmetrically) explain all correlations. This is the same reaction as for Bell's inequalities \cite{theo}, but there it is motivated by the fact that these inequalities refer to non-commutative measurements for which the assumption on several common causes is sensible. Here, this assumption is less acceptable, since we deal with commutative measurements. 

\comment{This paper is organized as follows. First, we briefly review PCC in the classical and quantum situations. The next section reminds the measurement formalism for identical particles. Then we show that the symmetric common cause is absent within the standard approach to joint probabilities for identical particles, which employs particle labeling. We work out an alternative approach towards joint probabilities, which explicitly avoids labeling the identical particles. We summarize in the last section. 
}

\section{The common cause principle in classical and quantum}
We are given two dependent random variables $A$ and $B$, with values $A=\{a\}_{1}^n$ and $B=\{b\}_{1}^n$. For simplicity we assume that $A$ and $B$ assume the same number of values. Now $C=\{c\}_1^m$ a common cause of $A$ and $B$, if 
\cite{reich,suppes,berko,szabo} 
\be
\label{1}
&& p_{AB|C}(a,b|c)=p_{A|C}(a|c)p_{B|C}(b|c),\\
&& p_{AB}(a,b)=\sum_{c=1}^m p_C(c)p_{A|C}(a|c)p_{B|C}(b|c), \\
&& p_C(c)>0, \qquad \sum_{c=1}^m p_C(c)=1. 
\label{1.1}
\ee
Hereafter, $C$ will denote the common cause holding (\ref{1.1}). We sometimes omit the lower indices for probabilities: $p_C(c)=p(c)$. 

Representation (\ref{1.1}) for the element-wise non-negative 
matrix $\{p_{AB}(a,b)\}_{a,b=1}^{n,n}$ is valid only when 
\be
m\geq {\rm rank}_+[\{p_{AB}(a,b)\} ],
\label{1.2}
\ee
where ${\rm rank}_+[\{p_{AB}(a,b)\} ]$ is the positive rank of matrix $\{p_{AB}(a,b)\} $ \cite{rank-nonnegative}. It relates to the usual rank \cite{rank-nonnegative}: 
\be
 {\rm rank}[\{p_{AB}(a,b)\} ]\leq {\rm rank}_+[\{p_{AB}(a,b)\} ]\leq n,
\label{1.22}
\ee
where the first inequality in (\ref{1.22}) follows from the definition of $ {\rm rank}[\{p_{AB}(a,b)\} ]$. 

For a given generic $\{p_{AB}(a,b)\}_{a,b=1}^{n,n}$, there are many different $C$'s that hold (\ref{1.1}, \ref{1.22}). To select a definite $C$ in (\ref{1}), one requires prior information, or one needs to apply inference rules such as the maximal entropy rule  \cite{hovh2023}.  
\comment{Ref.~\cite{gillis-review} reviews several specific classes of matrices $\{p_{AB}(a,b)\}_{a,b=1}^{n,n}$, for which (\ref{1.1}, \ref{1.22}) is unique at 
$n={\rm rank}_+[\{p_{AB}(a,b)\} ]$. However, these classes are not generic, i.e., they are not stable against perturbations.}  

Applications of (\ref{1}, \ref{1.1}) are usually supplemented by above conditions {\it (i)} and {\it (ii)}. Instead of {\it (i)} and {\it (ii)} we can apply another condition: given that $C$ is the common cause of $A$ and $B$, $A$ is not a common cause of $B$ and $C$, and $B$ is not a common cause of $A$ and $C$. These conditions allow to get rid of trivial screeners: 
Note that in (\ref{1}, \ref{1.1}) one can formally take $C=A$ or $C=B$. This
does not amount to any physically meaningful common cause: the causality is an irreflexive relation, and the event (or random variable) cannot cause itself. 

Following Reichanbach \cite{reich}, one can introduce in (\ref{1}) additional conditions for $p_{A|C}(a|c)$ and $p_{B|C}(b|c)$ \cite{mazzola1,mazzola2}. Such conditions can be meaningful in concrete applications, but we refrain from imposing them, since we are interested by general features of PCC. 

Eq.~(\ref{1}) relates to the Nonnegative Matrix Factorization (NMF), a basic method in data representation \cite{we_dsaa}. We mention this relation to stress that the purpose of NMF is that of PCC: to explain the dependencies in $A$ and $B$. However, NMF is more interested in approximate and generalizable representations, which are achieved for $n\ll {\rm rank}_+[\{p_{AB}(a,b)\} ]$ in (\ref{1}) \cite{we_dsaa}.

\comment{
This drawback is absent in approximate implementations of (\ref{1.1}), which are effective for $n\gg 1$ and $m\ll n$ \cite{we_dsaa}. Such approximation representations formalize the concept of efficient causes and relate to Nonnegative Matrix Factorization technique in machine learning \cite{we_dsaa}. In the present work, we restric ourselves with the exact implementation of (\ref{1.1}), where it follows from (\ref{1}).
}

Now consider two non-identical quantum subsystems $\A$ and $\B$. A non-entangled joint state is defined to have a density matrix $\rho_{\A\B}$:
\be
\label{3}
\rho_{\A\B}=\sum_{c=1}^m p_C(c)\rho_{\A}(c)\otimes\rho_{\B}(c),
\ee
where $\rho_{\A}(c)$ and $\rho_{\B}(c)$ are density matrices pertaining to different particles, $p_C(c)$ is a classical probability, and where $m$ is a natural number. In (\ref{3}), $C$ is a general common cause, because it explains all correlations generated by one-particle measurements described via projectors $P_k\otimes {Q}_l$, where $\sum_kP_k=1$ and $\sum_l { Q}_l=1$. $C$ is also a non-contextual common cause, since the conditional probabilities, e.g., $p(P_k=1|c)={\rm tr}(\rho_{\A}(c)P_k)$ depend only on the projector $P_k$ itself, and not on the full orthogonal set $\{P_k\}$, where $\sum_kP_k=1$. Certain entangled states can have contextual common causes; see e.g. \cite{werner1989}. 

For an entangled state $\rho$, the representation (\ref{3}) does not hold. 
As Bell's inequalities show \cite{bell_cuisine,wharton}, there are entangled states $\rho$, so that four joint probabilities generated by 4 non-commutative measurements (on particles described from $\rho$) do not have the same common cause $C$.

\section{Two identical particles}
Consider two identical quantum particles. The states of such particles live on a tensor-product Hilbert space ${\cal H}\otimes {\cal H}$; e.g., $\H$ incorporates both position and spin degrees of freedom. Quantum measurements for such a system are described by projectors which live in ${\cal H}\otimes {\cal H}$, and are permutation-symmetric. Since we aim to define and discuss probabilistic dependencies (correlations) between the two particles, we shall consider the two-particle measurements that employ single-particle features, i.e. single-particle detectors. Such features are defined via an orthogonal set of projections 
$\{P_k\}_{k=1}^n$ living in $\H$:
\BEA
\label{n1}
\sum_{k=1}^nP_k=1,\qquad P_kP_l=\delta_{kl}P_k.
\EEA
The two-particle measurement is described by symmetrized projectors which live in $\H\otimes \H$:
\begin{align}
&\{ P_{kk}\equiv P_k\otimes P_k\}_{k=1}^n , \quad \{
P_{kl}\equiv P_k\otimes P_l+P_l\otimes P_k\}_{k<l}^n, \nonumber\\
&\sum_{k=1}^nP_{kk}+\sum_{k<l}^nP_{kl}
=1\otimes 1, 
\label{n55}
\end{align}
where projectors are orthogonal and sum to one according to (\ref{n55}). 
For two identical particles living in ${\cal H}\otimes {\cal H}$, $P_k\otimes 1$ are not observable, e.g. because measuring and applying the Luders rule will violate the permutation symmetry of the post-measurement state. Only permutation-symmetric projectors are measurable for identical particles \footnote{This point is not universally accepted in the literature; see e.g. \cite{podgor,tiko1,tiko2,goyal}. If non-symmetric projectors are allowed for identical particles, one can directly start with the approach around (\ref{60}).}. 
To give a concrete example of $\{P_k\}_{k=1}^n$, let us assume 
$\H=\H_{\rm coordinate}\otimes \H_{\rm internal}$ with ${\rm dim}[\H_{\rm internal}]=2$:
\BEA
\label{fea}
\{P_k\}_{k=1}^5=\Big\{ \Pi_1\otimes \frac{1\pm\sigma_1}{2},
\Pi_2\otimes \frac{1\pm\sigma_3}{2}, \Pi_0\Big\},
\EEA
where $\sigma_1$ and $\sigma_3$ are the Pauli matrices, $\Pi_l$ ($l=0,1,2$) is the projector to the space-domain $\Omega_l$ with $\Pi_i\Pi_{l'}=\delta_{ll'}$ and  $\sum_{l=0}^2\Pi_l=1$, and where $\pm$ signs are chosen independently. 
Eq.~(\ref{fea}) refers to a set of single-particle features, where the particle is measured to be in space-domain $\Omega_1$ with spin $\sigma_1=\pm 1$ or in space-domain $\Omega_2$ with spin $\sigma_3=\pm 1$, or it is in $\Omega_0$. Hence, (\ref{fea}) describes two spin detectors located in $\Omega_1$ and $\Omega_2$, respectively. $\Omega_0$ refers to events where the particles are out of the detector range. 

Projectors in (\ref{n55}) refer to a symmetric hermitean operator 
$1\otimes {F}+{F}\otimes 1$ with eigenresolution ${F}=\sum_{k=1}^n f_kP_k$:
\BEA
\label{vano}
1\otimes {F}+{F}\otimes 1=2 \sum_{k=1}^nf_k P_{kk}
+\sum_{k<l=1}^n(f_k+f_l)P_{kl}.
\EEA
The derivation of (\ref{vano}) is facilitated by noting that the hermitean operator $P_k\otimes 1+1\otimes P_k$ has only two non-zero eigenvalues, $2$ and $1$ with (resp.) eigenprojectors $P_{kk}$ and $P_k\otimes 1+1\otimes P_k-P_k\otimes P_k$.

Generically (i.e. when there are not specific relations between $\{f_k\}_{k=1}^n$), the eigenvalues $\{2f_k\}_{k=1}^n$, $\{f_k+f_l\}_{k<l=1}^n$ of $1\otimes F+F\otimes 1$ are different. Hence, the projectors in (\ref{n55}) are measurable via measuring $1\otimes F+F\otimes 1$.

Based on measurement results generated by (\ref{n55}), there are at least two different ways for defining joint probabilities for two identical particles. They are discussed below.

\section{The standard approach to joint probabilities}
This intuitive approach for defining joint probabilities is aligned with Refs.~\cite{landau2013quantum,tiko1,tiko2}. Starting from (\ref{n55}) we assume that for the two identical particles there is joint probability $p(k,l)$ of two (independently defined random variables) with realizations $\{k\}_{k=1}^n$ and $\{l\}_{l=1}^n$, which is defined in the usual way:
\BEA
p(k,l)=p(l,k)={\rm tr}(\rho P_k\otimes P_l),
\label{n77}
\EEA
where we recall that $\rho$ is permutation symmetric.
Eq.~(\ref{n77}) does not imply that $P_k\otimes P_l$ for $k\not=l$ is observable, simply because $p(k,l)$ in (\ref{n77}) is easily recovered from the observed quantity: $p(k,l)= \frac{1}{2} {\rm tr}[\rho P_{kl}]$. The interpretation of (\ref{n77}) implicitly uses labeling of particles: the first (second) particle produces $k$ ($l$).

Let us return to the common cause formula (\ref{1}, \ref{1.1}) and apply it to the joint probability in (\ref{n77}). Once we deal with fundamentally identical particles, it is reasonable to assume that the common cause acts symmetrically on the two particle; i.e., instead of (\ref{1}, \ref{1.1}) we get
\begin{align}
\label{10}
&p(k,l|c)=p(k|c)p(l|c),\quad k,l=1,..n,\quad c=1,..,m,\\
& p(k,l)=\sum_{c=1}^m p(c)p(k|c)p(l|c),
\label{10.1}
\end{align}
where the symmetry of the common cause means that the same functions $\{p(...|c)\}_{c=1}^m$ apply to both random variables 
i.e., to $k$ and $l$. 

Denoting $\widetilde{p}(k,c)=\sqrt{p(c)}p(k|c)$ we can write (\ref{10.1}) in matrix notations as [$\widetilde{p}^{\rm T}$ is the transpose of $\widetilde{p}$]:
\BEA
\label{60}
p=\widetilde{p}\,\widetilde{p}^{\rm T}.
\EEA
Eq.~(\ref{60}) does not exist if the $n\times n$ symmetric
matrix $p$ is not positive semidefinite, e.g. if $p$ has a negative
eigenvalue. Now $n=2$ is the simplest situation when a negative eigenvalue of $p$ is possible. 

Eq.~(\ref{60}) can be invalid, even if all eigenvalues of $p$ are non-negative.
Positive (symmetric) semidefinite matrices that support (\ref{60}) with
$\widetilde{p}(k,c)\geq 0$ are called completely positive. The theory of such matrices is reviewed in \cite{gray}. For $n\geq 5$, there are entrywise nonnegative, positive semidefinite matrices that are not completely positive.

Thus, a symmetric joint probability (\ref{n77}) need not have a symmetric common
causes. This is not a problem for non-identical (hence distinguishable) particles. But for identical particles, a non-symmetric common cause means effective distinguishability. Before taking such a conclusion seriously, we need to look for a different approach. 

\section{Joint probabilities without implicit labels}
Probabilities produced via (\ref{n55}) admit another interpretation, where classical events are used instead of classical random variables. The advantage of this second approach, which was suggested in \cite{ghirardi1977some}, is that it does not involve labeling (numbering) of identical particles; hence, their inherent symmetry is manifest, and the supposed common cause accounts for this symmetry automatically. Define the following classical events
\BEA
\label{n8}
S_k=\{{\rm at~least~one~particle~has~feature}~ P_k=1\},\\
{\cal S}_{kk}=\{{\rm both~particles~have~feature}~ P_k=1\}.
\EEA 
Now we continue from (\ref{n55}) defining probabilities:
\BEA
\label{n9}
&& p(S_k\land S_l)={\rm tr}[\rho P_{kl}],\quad
p({\cal S}_{kk})={\rm tr}[\rho P_{kk}],\\
&&\sum_{k=1}^n p({\cal S}_{kk})+\sum_{k<l} p(S_k\land S_l)=1.
\label{normo}
\EEA
We can define from (\ref{n9}) probabilities of $p(S_k)$:
\BEA
\label{n10}
p(S_k)=p({\cal S}_{kk})+\sum_{l(\not=k)}p(S_k\land S_l)={\rm tr}[\rho Y_k],
\EEA
where $Y_k=P_k\otimes 1+1\otimes P_k-P_k\otimes P_k$ is a projector with obvious meaning. In particular, we note from (\ref{n9}, \ref{n10}) that $p(S_k\land S_l)={\rm tr}[\rho Y_kY_l]$, as it should be. Eq.~(\ref{n10}) implies a non-trivial relation:
\BEA
\label{n11}
p(S_1\lor...\lor S_n)=1,
\EEA
which is derived from (\ref{n55}), the general formula
$p(A\lor B)=p(A)+p(B)-p(A\land B)$, 
and $p(S_1\land ..\land S_\ell)={\rm tr}(\rho Y_1...Y_\ell)=0$
for $\ell\geq 3$. Likewise, we get from (\ref{n10}):
\BEA
\label{new}
{\cal S}_{kk}=S_k-\lor_{l(\not=k)}(S_k\land S_l).
\EEA

The dependency structure is now described by joint probabilities (\ref{n9}), where $\{S_k\}_{k=1}^n$ are events and not random variables. Recall that the formulation of the common cause principle by Reichenbach referred to correlations between events that are explained via random variables \cite{reich}. Hence, we want to find a screening random variable $C$ for all $n(n-1)/2$ (generically different) joint probabilities:
\BEA
\label{n12}
&& p(S_k\land S_l)=\sum_{c=1}^m p(c)p(S_k|c)p( S_l|c),\\ 
&& p(S_k)=\sum_{c=1}^m p(c)p(S_k|c), \quad k=1,...,n.
\label{n122}
\EEA
The advantage of (\ref{n12}, \ref{n122}), as compared to the approach suggested around (\ref{60}) is that no symmetry assumptions have to be made directly. 
Note that ${\cal S}_{kk}$ is not involved in (\ref{n12}) directly, since ${\cal S}_{kk}$ is not a joint probability of two separate events.

We continue with $n=3$, since this is a fully generic case; $n=2$ is not generic.
We work out $p({\cal S}_{11})$, $p({\cal S}_{22})$ and $p({\cal S}_{33})$ from 
(\ref{n10}, \ref{n122}) and put the results to the normalization condition (\ref{normo}). We find
\be
\label{omn}
&&0=\sum_{c=1}^m p(c)f\Big[p(S_1|c), p(S_2|c), p(S_3|c)  \Big], \\ 
&&f[x,y,z]\equiv 1+xy+xz+zy-x-y-z.
\ee
Now note that the minimum of $f[x,y,z]$ is achieved at the border of the allowed domain $0\leq x\leq 1$, $0\leq y\leq 1$, $0\leq z\leq 1$. To see this, one first shows that the only internal stationary point $(x,y,z)=(1/2,1/2,1/2)$ of $f[x,y,z]$ is a saddle. Taking (say) $z=0$, we again confirm that there are no minima internal to $0\leq x\leq 1$, $0\leq y\leq 1$. 

The minimal value is $f[x,y,z]=0$, and it is reached within 6 possible solutions: 3 solutions, where only one among $(x,y,z)$ is 1 and the others are 0. And 3 solutions, where only two among $(x,y,z)$ are 1 and the other is 0. Thus, we reach the following minimal screening structure for $m=3$:
\begin{align}
&p(c_1)=p(S_1\land S_2),\quad p(c_2)=p(S_1\land S_3),\nonumber\\ 
\label{ka1}
&p(c_3)=p(S_2\land S_3),\\
&p(c_4)=p({\cal S}_{11}),\quad p(c_5)=p({\cal S}_{22}),\quad p(c_6)=p({\cal S}_{33}).
\label{ka2}
\end{align}
Now $p(S_i|c_k)$ equals 1 or 0; e.g. $p(S_1|c_1)=p(S_2|c_1)=1$, $p(S_3|c_1)=0$. The structure of (\ref{ka1}, \ref{ka2}) is directly generalized for $n\geq 4$. 

We note, however, that (\ref{ka1}, \ref{ka2}) do not indicate any causality. 
To make this point clear, note that within (\ref{ka1}, \ref{ka2}) we do not have any further information about $C$. Hence, the screening events $c_k$ can be even expressed via $S_k$ and ${\cal S}_{kk}$; i.e., $c_1=S_1\land S_2$, $c_4={\cal S}_{11}$, {\it etc}. It is not clear why we should say in (\ref{ka1}) that $c_1$ caused $S_1$ and $S_2$, and not $S_1\land S_2$ caused $c_1$. Put differently, $C$ is a trivial screener, akin to examples we discussed before (\ref{3}). The same triviality conclusion applies to non-minimal common causes. Now a larger set of values of $C$ aggregates into $\{c_k\}_{k=1}^6$, which holds (\ref{ka1}, \ref{ka2}). 

\comment{
\begin{align}
\label{ka1}
&c_1=S_1\land S_2,\quad c_2=S_1\land S_3,\quad c_1=S_2\land S_3,\\
&c_4={\cal S}_{11},\quad c_5={\cal S}_{22},\quad c_6={\cal S}_{33}.
\label{ka2}
\end{align}
}

\section{Common causes for some dependencies}
Now we provide a realistic example showing that sensible common causes exist if we do not explain the dependencies in all joint probabilities $p(S_k\land S_l)$ in (\ref{n9}). Let us return to example (\ref{fea}), and make the standard assumption that the density matrices of the coordinates and the internal degrees of freedom are factorized \cite{landau2013quantum}, i.e., the full state $R$ of the two particle system reads
\BEA
\label{mu0}
R=\omega\otimes \rho,
\EEA
where $\omega$ and $\rho$ live (resp.) in ${\cal H}_{\rm coordinate}\otimes {\cal H}_{\rm coordinate}$ and
${\cal H}_{\rm internal}\otimes {\cal H}_{\rm internal}$, and where both $\omega$ and $\rho$ are permutation invariant. We also assume that only one (but certainly one) particle enters each coordinate detector:
\BEA
\label{mu1}
&& {\rm tr}\Big(\omega(\Pi_1\otimes \Pi_2+
\Pi_2\otimes \Pi_1\,)\Big)=1, \\
&& {\rm tr}\Big(\omega \,\Pi_1\otimes \Pi_2\Big)={\rm tr}\Big(\omega\, \Pi_2\otimes \Pi_1\Big)
=\frac{1}{2},
\label{mu2}
\EEA
where (\ref{mu2}) follows from (\ref{mu1}) upon noting that $\omega$ is permutation symmetric.

Using (\ref{mu0}, \ref{mu2}) we find that the joint probability of internal degrees of freedom reads
\BEA
\label{julius2}
p(\lambda,\tau)=
{\rm tr}\Big(\rho \,\frac{1+\lambda \sigma_1}{2}\otimes \frac{1+\tau \sigma_3}{2}\,\Big),
\EEA
where $\lambda=\pm 1$, $\tau=\pm 1$, and where
we again employed the fact that $\rho$ is permutation symmetric. Note that 
by analogy to (\ref{n9}, \ref{normo}), we can introduce events $S^{[0]}$ and $S^{[l]}_{\lambda}$ ($l=1,2$), and then (\ref{julius2}) is the joint probability ${p(S^{[1]}_{\lambda}\land S^{[2]}_{\tau} )}$ of two distinct classical events (or random variables). Now the upper index in $S^{[1]}_\lambda$ refers to the space detector, which serves as a distinguishing degree of freedom. Note that (\ref{julius2}) holds $p(1,-1)=p(-1,1)$.

The joint probability (\ref{julius2}) admits well-defined common causes $C$ \cite{hovh2023}. Among them, there are minimal-support, i.e., binary causes. If we do not know the concrete form of this cause, but know that it is binary, we can choose to work with the most likely binary common cause, which demonstrates nontrivial and interesting causality features \cite{hovh2023}. Those causes explain only certain dependencies, which do not nullify due to a special choice of the density matrices in (\ref{mu0}--\ref{mu2}). The assumptions in (\ref{mu0}--\ref{mu2}) are reasonable, but they are still non-generic: generically, we will need to explain all correlations, and this leads to trivial screeners, as we saw above. 

\section{Summary}
Can the joint probabilities generated by commutative measurements on two identical particles be explained via the principle of common cause (PCC)? There are two approaches for answering this question. They start from the same measurement data, but introduce different classical probability spaces. 

The first approach assumes implicit labels for the particles. Then the standard PCC applies, but for certain states it cannot be symmetric with respect to particles.
Allowing a nonsymmetric common cause means that at some (possibly hidden) level the identical particles can be distinguishable. Such an effective distinguishability for identical particles can be taken seriously; see \cite{we}. This goes beyond the standard quantum mechanics. 

The second approach avoids altogether introducing the particle labels. But now the screeners are trivial and do not refer to real causality. To avoid this conclusion, we can assume that correlations deduced within the commutative measurements on identical particles are not described via the same random cause variables: two or more such variables are needed. A particular case of this situation is seen above, where only non-zero correlations are explained via (sensible) common causes. This situation resembles the reaction to the non-existence of a single common cause variable for entangled states due to Bell's inequalities; see e.g. \cite{theo}. But there the main physical reason behind this reaction was that correlations that violate Bell's inequalities are deduced via incompatible (noncommutative) measurements; cf.~Ref.~\cite{gudder1984reality,allahverdyan2018excluding}. Here, the measurements are commutative, and assuming different common causes does not seem to be a well-founded approach. Another potentially useful option is to explain approximately all dependencies; cf.~\cite{we_dsaa}. We see that the existence of a common cause variable is problematic not only due to noncommutative measurements, as witnessed by Bell's inequalities, but also for identical quantum particles under commutative measurements.  

\acknowledgments

We thank Arik Avagyan and Vahagn Abgaryan for their interest in this work. The HESC of Armenia supported us via grants 24FP-1F030 and 21AG-1C038. 

\bibliographystyle{IEEEtran}
\bibliography{paper}

\section{Appendix: Common cause, quantum entanglement and Bell inequalities}
The assumptions that go into derivations of Bell's inequalities are frequently confusing; see Refs.~\cite{szabo,berko,wharton}. The proper Bell inequalities are not the focus of the present paper. But a close relation between the principle of the common cause (PCC) and Bell's inequalities led us to rethink and, below, present several interrelated concepts in a compact form. 

We start the most transparent approach to Bell's inequalities. Assume that quantum observables $X$ with eigenvalues $\{x\}$ and $Y$ with eigenvalues $\{y\}$ commute and hence can be measured together. Let $p_{XY}(x,y)$ be the corresponding joint probabilities, while $\sum_y p_{XY}(x,y)=p_X(x)$ and $\sum_x p_{XY}(x,y)=p_Y(y)$ are the marginal probabilities. Note that $X$ in $p_{X}(x)$ denotes the random variable $X$ to which $x$ belongs. If $C$ is the common cause of $X$ and $Y$, we have [cf.~(\ref{1})]
\BEA
\label{osaka1}
p_{XY}(x,y)=\sum_cp(c)p_X(x|c)p_Y(y|c).
\EEA
Consider two distinguishable particles and define the following four observables:
\BEA
A_1=\sigma_{z}\otimes 1, \quad
B_1=1 \otimes \sigma_{z}, \nonumber\\
A_2=\sigma_{x}\otimes 1, \quad
B_2=1 \otimes \sigma_{x}.
\label{h55}
\EEA 
Note that $[A_1,B_1]=[A_2,B_2]=[A_1,B_2]=[A_2,B_1]=0$ and hence the corresponding pairs can be measured together and produce valid joint 
probabilities from quantum mechanics. 
Next, we check whether all four joint probabilities share the same common cause variable $C=\{c\}$: (\ref{osaka1}) holds with the same $C=\{c\}$ and $p(c)$ for
\BEA
\label{osaka2}
(X,Y)=(A_1,B_1),\,(A_2,B_2),\,(A_1,B_2),\,(A_2,B_1).
\EEA
Sample $C$ with $p(c)$ and generate an outcome
$c_\alpha$. Sample independently $p_{A_i}(a_i|c_\alpha)$ and $p_{B_i}(b_i|c_\alpha)$ ($i=1,2$) and generate outcomes 
$a_{i\alpha}=\pm 1$ and $b_{i\alpha}=\pm 1$ \cite{eberhard}.
Consider:
\BEA
\label{h10}
(a_{1\alpha}+a_{2\alpha})b_{1\alpha}+(a_{1\alpha}-a_{2\alpha})b_{2\alpha}.
\EEA
Since $a_{i\alpha}=\pm 1$, only one in term 
(\ref{h10}) is non-zero. Summing (\ref{h10}) over 
sufficiently many samples we find 
\BEA
\label{h12}
|\langle A_1B_1\rangle+\langle A_2B_1\rangle+\langle A_1B_2\rangle-\langle A_2B_2\rangle|\leq 2.
\EEA
This is the Bell-CHSH inequality \cite{bell_cuisine,fraassen,wharton,szabo}, and it contains only quantities that can be calculated from quantum mechanics, because (\ref{h12}) involves only joint probabilities of commuting quantum observables. This however does not mean that the averages involved in (\ref{h12}) can be
calculated via a single measurement, because there are still non-commuting observables: $[A_1,B_1]\not =0$ and $[A_2,B_2]\not =0$.
It is well-known that (\ref{h12}) is violated for certain entangled quantum states 
\cite{bell_cuisine,fraassen,wharton,szabo}.

Within this formulation, (\ref{osaka1}, \ref{osaka2}) is an attempt to find a general cause for joint probabilities generated via different (noncommuting) measurements, and the violation of Bell's tells that this attempt fails due to noncommutativity for certain quantum states. The locality can be emphasized only in the context of applying PCC itself: once the particles are well-separated, they do not influence each other, and PCC can be employed; cf.~{\it (i)} in the introduction. Recall that (\ref{h55}) implies non-identical particles. However, Bell's inequalities apply also to identical particles provided we employ the reasoning around (\ref{n77}).

The main starting point of the second approach is that the choice of the measurement and the measurement results are assumed to be different random variables; see e.g. \cite{spekkens}. The random variable $A$ $[B]$ with realizations $(A_1,A_2)$ [$(B_1,B_2)$] denotes the measurement done on the first [second] particle; cf.~(\ref{h55}). An essential assumption is that the operators $A_1$ and $A_2$ [$B_1$ and $B_2$] have the same eigenvalues. Then the measurement results can be introduced via two other random variables $U=\{u\}$ [$V=\{v\}$], where $\{u\}$ [$\{v\}$] denotes the eigenvalues of $A_1$ and $A_2$ [$B_1$ and $B_2$]. Thus, we have 4 random variables $A$, $B$, $U$, and $V$. The Born probabilities refer to $p(u,v|A_k,B_l)$. One can (but need not) conduct experiments with them $A$ and $B$ as random variables. 

\comment{The operational part of quantum mechanics amounts to selecting a
measurement $A$ (we assume that the measurements are labeled by a
discrete or a continuous index $A$) and recording the probabilities
$p(a|A)$ that refer to eigenvalues of the measured observable $A$. For
commuting observables one can introduce joint probabilities
$p(a,b|A,B)$. We do not mean that e.g. $A$ itself has a known probability. }

The no-signalling condition amounts to
($k,l=1,2$)
\begin{align}
&{\sum}_v p(u,v|A_k,B_l)=p(u|A_k),\\
&{\sum}_u p(u,v|A_k,B_l)=p(v|B_l).
\label{ns}
\end{align}
Within the previous approach, the no-signalling was just a triviality needed for consistent definition of joint probabilities. Eqs.~(\ref{ns}) are nontrivial, and they 
prevent superluminar communication between spatially separated measurements.

The common cause principle leads to looking at a random variable
$C=\{c\}$, which is located in the common past of $U$ and $V$, 
and factorizes (screens) the dependencies present in $p(a,b|A,B)$:
\BEA
\label{hh1}
p(u,v|A_k,B_l,c)=p(u|A_k,B_l,c)p(v|A_k,B_l,c),
\EEA
which implies
\BEA
&p(u,v|A_k,B_l)=\sum_c p(u,v,c|A_k,B_l)\nonumber\\
&=\sum_c p(c|A_k,B_l) p(a|A_k,B_l,c)p(b|A_k,B_l,c).
\label{h3}
\EEA
The first assumption imposed on (\ref{h3}) is that $C$ does not depend on $A$ and $B$:
\BEA
\label{h4}
p(c|A_k,B_l)=p(c).
\EEA
The second assumption in (\ref{h3}) is 
\BEA
&&p(u|A_k,B_l,c)=p(u|A_k,c), \nonumber\\ 
&&p(v|A_k,B_l,c)=p(v|B_l,c). 
\label{h2}
\EEA
Conditions (\ref{ns}) can be derived via (\ref{h2}, \ref{h4}). Indeed, (\ref{h4}) implies 
$p(c|A_k,B_l)=p(c)=p(c|A_k)=p(c|B_l)$. Then (\ref{ns}) is deduced from (\ref{h2}). However, the inverse derivation is generally impossible, i.e., (\ref{h2}) does not follow from (\ref{ns}). 

Thus, (\ref{h2}) are the generalized locality conditions that are stronger than no-signaling conditions (\ref{ns}). Eqs.~(\ref{h4}, \ref{h2}) lead from (\ref{h3}):
\BEA
\label{h5}
p(u,v|A_k,B_l)=\sum_c p(c) p(u|A_k,c)p(v|B_l,c).
\EEA
Eq.~(\ref{h5}) is analogous to (\ref{osaka1}, \ref{osaka2}) and allows to derive the Bell-CHSH inequality (\ref{h12}).

\end{document}